\documentclass[a4paper,11pt]{amsart}
\usepackage{graphicx}
\usepackage{amssymb}
\begin{document}

\hyphenation{gra-vi-ta-tio-nal re-la-ti-vi-ty Gaus-sian
re-fe-ren-ce re-la-ti-ve gra-vi-ta-tion Schwarz-schild
ac-cor-dingly gra-vi-ta-tio-nal-ly re-la-ti-vi-stic pro-du-cing
de-ri-va-ti-ve ge-ne-ral ex-pli-citly des-cri-bed ma-the-ma-ti-cal
de-si-gnan-do-si coe-ren-za pro-blem gra-vi-ta-ting geo-de-sic
per-ga-mon cos-mo-lo-gi-cal gra-vity cor-res-pon-ding
de-fi-ni-tion phy-si-ka-li-schen ma-the-ma-ti-sches ge-ra-de
Sze-keres con-si-de-red tra-vel-ling ma-ni-fold re-fe-ren-ces
geo-me-tri-cal in-su-pe-rable sup-po-sedly at-tri-bu-table
Bild-raum in-fi-ni-tely counter-ba-lan-ces iso-tro-pi-cally
pseudo-Rieman-nian cha-rac-te-ristic geo-de-sics
Koordinaten-sy-stems ne-ces-sary Col-la-bo-ra-tion}

\title[On the GR-speeds of light and particles] {{\bf On the GR-speeds of light and particles\\(\emph{A remembrance of things past)}}}

\author[Angelo Loinger]{Angelo Loinger}
\address{A.L. -- Dipartimento di Fisica, Universit\`a di Milano, Via
Celoria, 16 - 20133 Milano (Italy)}
\author[Tiziana Marsico]{Tiziana Marsico}
\address{T.M. -- Liceo Classico ``G. Berchet'', Via della Commenda, 26 - 20122 Milano (Italy)}
\email{angelo.loinger@mi.infn.it} \email{martiz64@libero.it}

\vskip0.50cm

\begin{abstract}
There are no limits for the speeds of light and particles in
general relativity (GR). Four examples illustrate this basic
result, which is too often neglected.
\end{abstract}

\maketitle

\vskip0.80cm \noindent \small PACS 04.20 -- General relativity.
\normalsize

\vskip1.20cm \noindent \textbf{1.} -- The recent interesting
experiments by the OPERA Collaboration concerning the neutrino
velocity \cite{1} have given rise indirectly to renewed
discussions about the velocities of light and particles in general
relativity. -- In this Note we emphasize the following fundamental
concepts: \emph{i}) In GR the light velocity can have \emph{any}
value between zero and infinite, according to the values of the
metric tensor; \emph{ii}) \emph{Idem} for the particle velocity,
which however must always be lower than the light velocity in the
\emph{same} pseudo-Riemannian manifold, but it can be greater than
$c$ (the light velocity \emph{in vacuo}, \emph{i.e.} in the
absence of any field) in some instances.

\par We shall illustrate these statements with four simple
examples.

\vskip1.20cm \noindent \textbf{2.} -- A preliminary consideration.
In an axiomatic treatment of GR -- see, \emph{e.g.}, Hilbert
\cite{2} -- the theory is divided in two distinct parts. The first
part is devoted to the properties of the Einsteinian field
equations, without any indication of the Riemannian or
pseudo-Riemannian character of the interested manifolds.

\par In the second part, according to the \emph{Erfahrung},
Hilbert specifies the nature of the interval $\textrm{d}s^{2}$,
which can be time-like, light-like, or space-like. The spacetime
manifolds of physical interest must have a well-defined
pseudo-Riemannian character.

\par The points \emph{i}) and \emph{ii}) of sect.\textbf{1} are
implicitly contained in this axiom, \emph{i.e.} they are a logical
consequence of it.

\vskip1.20cm \noindent\textbf{3.} -- Let us consider the
Einsteinian stationary field which exists in a reference frame
uniformly rotating \cite{3}, \cite{4}. The interval
$\textrm{d}s^{2}$ is:

\begin{equation} \label{eq:one}
\textrm{d}s^{2} = \left(c^{2} -  \Omega^{2}r^{2}\right)
\textrm{d}t^{2} - 2 \, \Omega \, r^{2} \, \textrm{d}\varphi \,
\textrm{d}t - \textrm{d}\sigma^{2} \quad,
\quad(c^{2}>\Omega^{2}r^{2}) \quad,
\end{equation}

\begin{equation} \label{eq:oneprime}
\textrm{d}\sigma^{2} \equiv \textrm{d}r^{2} +
r^{2}\textrm{d}\varphi^{2} + \textrm{d}z^{2}\tag{1$'$} \quad,
\end{equation}

where $\Omega$ is the angular speed of rotation, and $r, \varphi,
z$ are cylindrical space-coordinates. Light motions are
characterized by $\textrm{d}s^{2}=0$. If $\Omega$ is \emph{small},
we can keep only the terms of \emph{first} order in $\Omega \,
r^{2}/c$. We get:

\begin{equation} \label{eq:two}
\textrm{d}t = \frac{\textrm{d}\sigma}{c} \pm \frac{2\Omega}{c^{2}}
\left(\frac{1}{2} \, r^{2}  \textrm{d}\varphi \right) \quad,
\end{equation}

where the double sign $(\pm)$ corresponds to the two directions of
the light-ray motion.

\par If we put $r=a_{1}$, the Earth's equatorial radius $(a_{1}=
6,378,137 \, \textrm{m})$, and $\Omega = \omega_{\textrm{E}}$, the
terrestrial rotation speed $(\omega_{\textrm{E}} = 7.2921151467
\times 10^{-5} \, \textrm{rad/s})$, for a full turn (within an
optical fibre) of a light-ray around the equator we have:

\begin{equation} \label{eq:three}
\Delta t = \frac{2\pi a_{1}}{c} \pm
\frac{2\omega_{\textrm{E}}}{c^{2}} \, \pi a_{1}^{2} \quad;
\end{equation}

since $\pi a_{1}^{2}= 1.27802 \times 10^{14} \, \textrm{m}^{2}$,
and $2\omega_{\textrm{E}}/c^{2}=1.6227 \times 10^{-21} \,
\textrm{m}^{-2}\textrm{s}$, the second term at the right-side of
(\ref{eq:three}) -- which gives a so-called Sagnac effect -- is
equal to $207.4 \, \textrm{ns}$.

\par The light velocity $w$ is:

\begin{equation} \label{eq:four}
w = \frac{2\pi a_{1}}{\Delta t} = c \mp \omega_{\textrm{E}} \,
a_{1} \quad ; \quad (\omega_{\textrm{E}} \, a_{1} = 0.4651011 \,
\textrm{km/s})\quad;
\end{equation}

we see that if the motions of the light-ray and of the reference
frame happen in opposite directions, the light velocity is
\emph{greater} than $c$.

\par It is useful to compute $w$ as an instantaneous speed. From
$\textrm{d}s^{2}=0$, we obtain:

\begin{equation} \label{eq:five}
\frac{\textrm{d}t}{\textrm{d}\varphi} = \pm
\frac{\omega_{\textrm{E}} \, a_{1}^{2}}{c^{2}} + \left(
\frac{\omega_{\textrm{E}}^{2} \,a_{1}^{4}}{c^{4}} +
\frac{a_{1}^{2}}{c^{2}}\right)^{1/2} \approx \,  \pm
\frac{\omega_{\textrm{E}} \, a_{1}^{2}}{c^{2}} + \frac{a_{1}}{c}
\quad ;
\end{equation}

from which:

\begin{equation} \label{eq:six}
a_{1}  \frac{\textrm{d}\varphi}{\textrm{d}t} \approx
\frac{c^{2}}{c\pm \omega_{\textrm{E}} \, a_{1}} \quad ;
\end{equation}

eq. (\ref{eq:six}) coincides with eq. (\ref{eq:four}) if we
neglect $\omega_{\textrm{E}}^{2} \, a_{1}^{2}$.

\vskip1.20cm \noindent\textbf{4.} -- With a good approximation,
the Einsteinian metric near the Earth can be written as follows
(see, \emph{e.g.}, \cite{4}):

\begin{equation} \label{eq:seven}
\textrm{d}s^{2} \approx \left( 1+ \frac{2V}{c^{2}}\right) c^{2}
\textrm{d}t^{2} - 2 \omega_{\textrm{E}} \, r^{2} \sin^{2}\vartheta
\, \textrm{d}\varphi \textrm{d}t - \left( 1-
\frac{2V}{c^{2}}\right) \textrm{d}\sigma^{2} \quad,
\end{equation}

\begin{equation} \label{eq:sevenprime}
\textrm{d}\sigma^{2} \equiv \textrm{d}r^{2} + r^{2}
\left(\textrm{d}\vartheta^{2} + \sin^{2}\vartheta \,
\textrm{d}\varphi^{2}\right) \tag{7$'$} \quad,
\end{equation}

if we use isotropic polar coordinates $r, \vartheta, \varphi$.

\begin{equation}\label{eq:eigth}
V \approx - \frac{GM_{\textrm{E}}}{r} \left[ 1- \mathcal{J}_{2} \,
\Big(\frac{a_{1}}{r}\Big)^{2} P_{2} (\cos\vartheta) \right] \quad
,
\end{equation}

where: $M_{\textrm{E}}$ is the Earth's mass, $\mathcal{J}_{2}$ is
Earth's quadrupole moment coefficient, $P_{2}$ is the Legendre
polynomial of degree $2$; $(GM_{\textrm{E}} = 3.986004418 \times
10^{14} \, \textrm{m}^{3}\textrm{s}^{-2}$ ; $\mathcal{J}_{2}=
1.0826300 \times 10^{-3}$). From $\textrm{d}s^{2}=0$, we have,
keeping only the terms of \emph{first} order in
$\omega_{\textrm{E}} \, r^{2}/c$ and $V/c^{2}$:

\begin{equation} \label{eq:nine}
c \, \textrm{d}t \approx \frac{\textrm{d}\sigma \pm
\omega_{\textrm{E}} \, r^{2} \sin^{2}\vartheta \,
\textrm{d}\varphi / c}{1+2V/c^{2}} \quad ;
\end{equation}

for a full turn of a light-ray around the equator, we get
$[P_{2}(\cos \pi /2) = -1/2]$:

\begin{equation} \label{eq:ten}
a_{1} \frac{\textrm{d}\varphi}{\textrm{d}t} \approx
\frac{c^{2}}{c\pm \omega_{\textrm{E}} \, a_{1}} \left[ 1 -
\frac{2|V(a_{1})|}{c^{2}} \right] \quad,
\end{equation}

where

\begin{equation} \label{eq:eleven}
|V(a_{1})|  \approx \frac{GM_{\textrm{E}}}{a_{1}} (1+0.54 \times
10^{-3}) \approx \frac{GM_{\textrm{E}}}{a_{1}} \quad,
\end{equation}

from which:

\begin{eqnarray}\label{eq:twelve}
a_{1} \frac{\textrm{d}\varphi}{\textrm{d}t}  & \approx &
\frac{c^{2}}{c\pm \omega_{\textrm{E}} \, a_{1}} \left( 1-
\frac{2GM_{\textrm{E}}/c^{2}}{a_{1}}\right) \approx
\nonumber\\
& & \approx (c \mp \omega_{\textrm{E}} \, a_{1}) \times (1-1.4
\times 10^{-9}) \approx
\nonumber\\
& & \approx (c \mp 0.4651011) \, \textrm{km/s} \quad .
\end{eqnarray}

A result which is essentially coincident with that of the mere
rotation.

\vskip1.20cm \noindent\textbf{5.} -- In Newton's theory, Laplace
equation $\nabla^{2} U=0$ admits the two solutions $U = \pm GM/r $
for a material point. Analogous situation in GR. Let us start from
the standard (Hilbert-Droste-Weyl) form of solution to
Schwarzschild's problem of the gravitational field created by a
point-mass $M$:

\begin{equation} \label{eq:thirteen}
\textrm{d}s^{2} = \left( 1 - \frac{2m}{r}\right) c^{2}
\textrm{d}t^{2} -  \left( 1 - \frac{2m}{r}\right)^{-1}
\textrm{d}r^{2} -  r^{2} (\textrm{d}\vartheta^{2} +
\sin^{2}\vartheta \, \textrm{d}\varphi^{2}) \quad ,
\end{equation}

where $m \equiv GM/c^{2}$ is an integration constant, that is
specified in this way because we wish to have the Newton solution
at great values of $r$. Now, the Einstein equations $R_{jk}=0$,
$(j,k=0,1,2,3)$, admit also the following solution, which has a
mere mathematical meaning (even if $\textrm{d}s^{2} \geq 0$):

\begin{equation} \label{eq:fourteen}
\textrm{d}s^{2} = \left( 1 + \frac{k}{r}\right) c^{2}
\textrm{d}t^{2} -  \left( 1 + \frac{k}{r}\right)^{-1}
\textrm{d}r^{2} -  r^{2} (\textrm{d}\vartheta^{2} +
\sin^{2}\vartheta \, \textrm{d}\varphi^{2}) \quad ,
\end{equation}

where $k>0$ is an integration constant; the condition
$\textrm{d}s^{2} = 0$ implies:

\begin{equation} \label{eq:fifteen}
\left| \frac{\textrm{d}r}{\textrm{d}t}\right|_{\textrm{LIGHT}} = c
\left(1+\frac{k}{r} \right)  > c \quad ;
\end{equation}

and for the radial geodesic of a test-particle we have:

\begin{equation} \label{eq:sixteen}
\left| \frac{\textrm{d}r}{\textrm{d}t}\right|_{\textrm{PARTICLE}}
= c \left(1+\frac{k}{r} \right) \left[ 1 - |A| \Big(1+
\frac{k}{r}\Big) \right] \quad ,
\end{equation}

where $A$ is an integration constant (for $A=0$, we re-obtain eq.
(\ref{eq:fifteen})). If $|A|$ is sufficiently small, \emph{the
particle velocity is greater than} $c$ for all the values of $r$
greater than a suitable value.

\vskip1.20cm \noindent\textbf{6.} -- The metric of Friedmann's
model of universe corresponding to a positive space curvature is
represented by the following interval $\textrm{d}s^{2}$ (where $x,
y, z, r$ are dimensionless coordinates which give labels to the
galaxies):

\begin{equation} \label{eq:seventeen}
\textrm{d}s^{2} = c^{2}\textrm{d}t^{2} - A^{2}(r^{2}) \, F^{2}(t)
\, \textrm{d}\sigma^{2} \quad; \quad (r^{2}=x^{2}+y^{2}+z^{2})
\quad ,
\end{equation}

\begin{equation} \label{eq:seventeenprime}
\textrm{d}\sigma^{2} =
\textrm{d}x^{2}+\textrm{d}y^{2}+\textrm{d}z^{2} \tag{17$'$} \quad
,
\end{equation}

\begin{equation} \label{eq:seventeensecond}
A(r^{2}) = 1 + (r^{2}/4) \tag{17$''$} \quad .
\end{equation}

The function $F(t)$ -- \emph{scale factor} -- is determined by
Einstein field equations (with a ``dust-like'' matter tensor), and
is a usual cycloid. Thus, we have a cosmic periodic oscillation
between $F=0$ and $F=F_{\textrm{max}}$.

\par From $\textrm{d}s^{2} =0$, we get:

\begin{equation} \label{eq:eighteen}
\left|
\frac{\textrm{d}\sigma}{\textrm{d}t}\right|_{\textrm{LIGHT}} =
\frac{c}{A(r^{2}) \, F(t)}  \quad ;
\end{equation}

we see that the light velocity is infinite at every big-bang and
at every big-crunch. --

\vskip1.80cm \emph{Acknowledgment}. We are very grateful to Dr. L.
Stanco of the OPERA Collaboration for the epistolary discussions
about several passages of the papers quoted in \cite{1}.

\end{document}